\documentclass[11pt]{article}
\usepackage[margin=1in]{geometry}
\usepackage{graphicx}
\usepackage{float}
\usepackage{verbatim}

\def\beq{\begin{equation}}
\def\eeq#1{\label{#1}\end{equation}}
\def\eeqn{\end{equation}}

\def\beqa{\begin{eqnarray}}
\def\eeqa#1{\label{#1}\end{eqnarray}}
\def\eeqan{\end{eqnarray}}

\let\bar=\overbar

\def\Dslash{\not{\hbox{\kern-4pt $D$}}}
\def\dslash{\not{\hbox{\kern-2pt $\del$}}}

\def\msb{{\bar{\ssstyle M \kern -1pt S}}}

\def\Title#1{\begin{center} {\Large {\bf #1} } \end{center}}
\def\Author#1{\begin{center} {\normalsize {\sc #1} } \end{center}}
\def\Institution#1{\begin{center} {\normalsize {\it #1} } \end{center}}
\def\Abstract#1{\noindent {\normalsize {\bf Abstract:} {\normalfont #1}}}
\def\Conference{\vspace{4mm}\begin{raggedright} {\normalsize {\it Talk presented at the 2019 Meeting of the Division of Particles and Fields of the American Physical Society (DPF2019), July 29--August 2, 2019, Northeastern University, Boston, C1907293.} } \end{raggedright}\vspace{4mm}}

\begin{document}

\Title{Optimizing the Performance of the CMS ECAL Trigger for Runs 2 and 3 of the CERN LHC}

\Author{Abraham Tishelman-Charny}

\Institution{Department of Physics\\ Northeastern University, Boston MA}

\begin{center}
    \textit{On behalf of the CMS Collaboration}
\end{center}

\Abstract{
The CMS Electromagnetic Calorimeter (ECAL) is a high resolution crystal calorimeter operating at the CERN LHC. It is responsible for the identification and precise reconstruction of electrons and photons in CMS, which were crucial in the discovery and subsequent characterization of the Higgs boson. It also contributes to the reconstruction of tau leptons, jets, and calorimeter energy sums, which are vital components of many CMS physics analyses. \\ \newline
The ECAL trigger system employs fast digital signal processing algorithms to precisely measure the energy and timing information of ECAL energy deposits recorded during LHC collisions. These trigger primitives are transmitted to the Level-1 trigger system at the LHC collisions rate of 40 MHz. These energy deposits are then combined with information from other CMS sub-detectors to determine whether the event should trigger the readout of the data from CMS to permanent storage. \\ \newline
This presentation will summarize the ECAL trigger performance achieved during LHC Run 2 (2015-2018). It will describe the methods that are used to provide frequent calibrations of the ECAL trigger primitives during LHC operation. These are needed to account for radiation-induced changes in crystal and photodetector response, and to minimize the spurious triggering on direct signals in the photodetectors used in the barrel region ($\eta<$1.48). Both of these effects are increased relative to LHC Run 1 (2009-2012), due to the higher luminosities experienced in Run 2. \\ \newline
Further improvements in the energy and time reconstruction of the CMS ECAL trigger primitives are being explored for LHC Run 3 (2021-23), using additional features implemented in the on-detector readout. These are particularly focused on improving the performance at the highest instantaneous luminosities (which will reach or exceed 2$\times 10^{34}\textrm{cm}^{-2}\textrm{s}^{-1}$ in Run 3) and in the most forward regions of the calorimeter ($\eta >$2.5), where the effects of detector aging will be the greatest. The main features of these improved algorithms will be described and preliminary estimates of the potential performance gains will be given.}

\Conference 
\thispagestyle{empty}
\newpage

\section{Introduction}

The main purpose of the electromagnetic calorimeter (ECAL) \cite{ECAL_TDR} of the Compact Muon Solenoid (CMS) \cite{CMS} experiment at the CERN LHC is to measure the energies of electrons and photons with high precision and accuracy. The ECAL is composed of 75,848 lead tungstate (PbW$\mathrm{O_{4}}$) crystals organized into a barrel (EB) section ($|\eta|<1.48$) and two endcaps (EE), which extend the coverage to $|\eta|=3$. Scintillation light from the crystals is recorded by APD photodetectors in EB and VPT photodetectors in EE, and is amplified, sampled, and digitized by on-detector electronics every 25ns to form signal waveforms whose amplitudes correspond to the energy deposits in each crystal. Because of the finite size of the on-detector data buffers, CMS needs to quickly determine if the full information from an event is worth keeping for further processing, so that it can decide in time if the on-detector buffer space can be used for information from the next event \cite{ECAL_Trigger}. \\ \newline 
Because of the high rate of collisions in CMS, it is not possible to save information from all events and CMS must trigger only on ``interesting" physics events. This is achieved via a two-stage triggering system. The Level-1 trigger (L1), implemented in custom hardware, combines coarse-grained information from various detector subsystems to reduce the event rate to below 100 kHz. The second level (High Level Trigger or HLT) is implemented in software and uses fine-grained detector information to further reduce the event rate to $\approx$ 1kHz. To help accomplish this, ECAL produces trigger primitives (TPs) every 25 ns that are sent to the Level-1 trigger. These TPs comprise a transverse energy measurement for a matrix of crystals, one or two feature bits, and a bunch crossing (BX) assignment. One of these feature bits, the strip fine grain veto bit (sFGVB), is used to reject anomalous signals in the barrel APD photodetectors \cite{sFGVB}. The ECAL TPs are paired with other subdetector information to form electron, photon, jet, and transverse energy sums and, if these are sufficiently energetic to satisfy the Level-1 trigger requirements, the event is considered interesting and the data buffers are read out and sent to the HLT for further processing \cite{ECAL_DAQ}. \\ \newline 
The signal amplitude for each crystal is estimated from the digitized signal pulses by the application of FIR (Finite Impulse Response) weights to five samples around the signal peak \cite{ECAL_Reco}. For a given sample, the estimated signal amplitude contribution is defined as the sample value in analog-to-digital-converter (ADC) counts multiplied by the weight assigned to that sample. The weights are derived from measured EB and EE pulses recorded in electron test beams

\section{Trigger in Run 2}
During Run 2 of the LHC (2015-2018), a greater instantaneous luminosity, in addition to a greater integrated luminosity, was reached compared to Run 1 (2011-2012). This resulted in larger radiation-induced effects that must be mitigated to preserve the trigger performance. \\ \newline 
Multiple time dependent corrections were made during Run 2 in order to maintain a stable TP energy scale, maintain the L1 spike killer performance for anomalous signals in EB, and minimize TPG pre-firing. Some of these methods include crystal inter-calibration, laser response corrections, pedestal drift corrections, and pulse shape corrections.

\subsection{Crystal Intercalibration}
Crystal intercalibration is needed to equalize the response of the individual crystals and is used to maintain a stable energy scale for TPs. This was updated twice in 2018 to account for radiation-induced changes in response. Each ECAL crystal is assigned an intercalibration constant derived from physics events, including $\pi^{0}\rightarrow\gamma\gamma$, the $\phi$-symmetry of minimum bias events, and electrons from Z and W boson decays \cite{Calibration}. 



\subsection{Laser Corrections}
The second method for maintaining a stable TP energy scale are time-dependent laser response corrections \cite{Calibration}. As ECAL receives more radiation from LHC collisions, the crystal transparency is decreased, leading to a decreased light output recorded by the ECAL electronics for a given signal amplitude. In order to account for this, a laser is shone onto all crystals and a measurement of their light yield is taken \cite{ECAL_Laser}, seen in Fig 1:

\begin{figure}[H]
\centering
\includegraphics[height=4in]{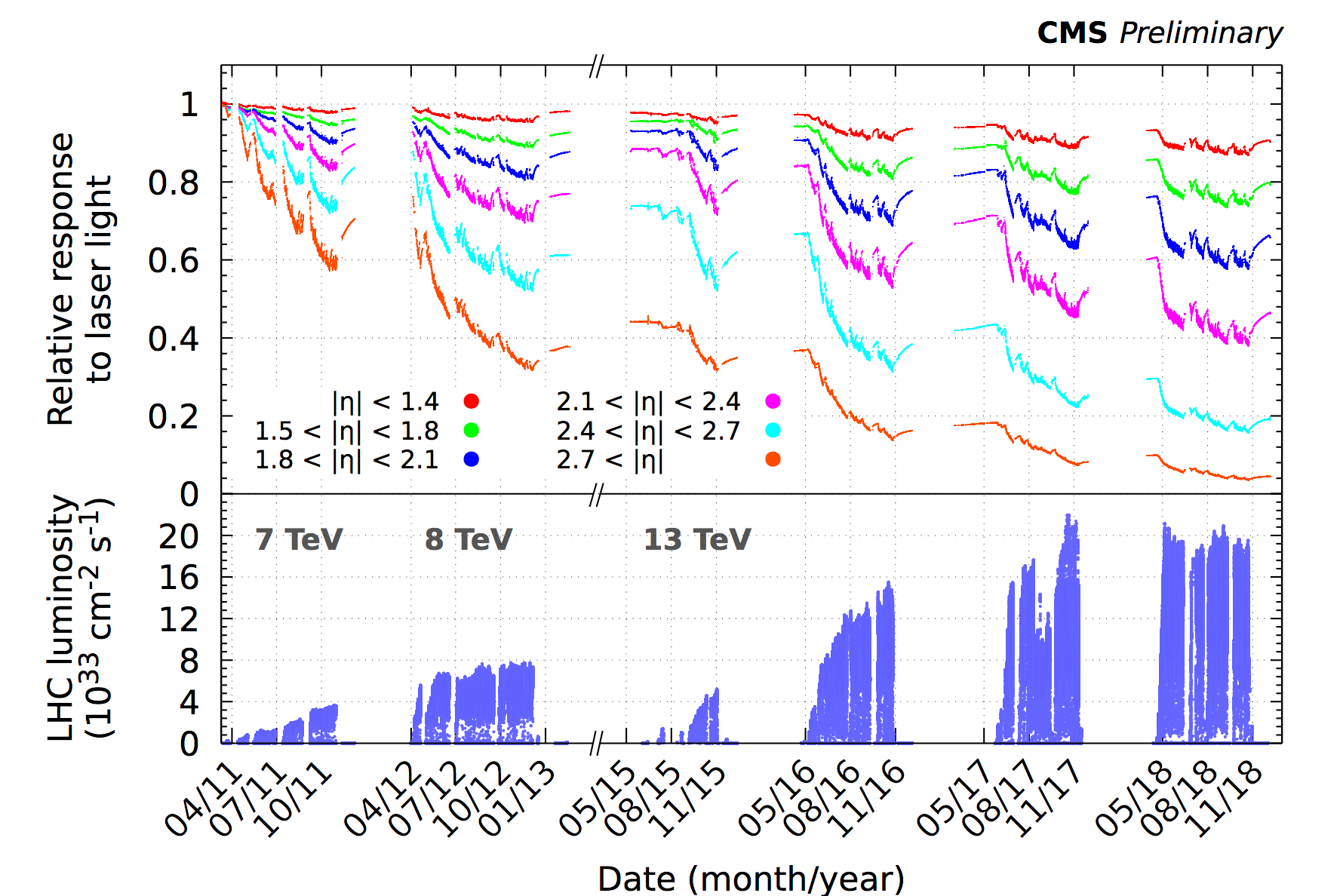}
\caption{Relative ECAL laser response versus time in LHC Runs 1 and 2, for several bins of pseudorapidity, $\eta$. The lower section of the figure displays the LHC luminosity over time. As more luminosity is delivered to CMS, relative response to laser light decreases.}
\end{figure}

\noindent The lower the light yield, the larger a correction factor is applied, to normalize the crystal response. It is noticeable that crystal light response decreases as $\eta$ increases, as these crystals receive more radiation. In 2018, this correction was updated at L1 and HLT twice per week, and was applied 52 times in total. 


\subsection{Pedestal Drift Corrections}
At L1, the sFGVB is used to determine if a signal is anomalous or not \cite{sFGVB}. In EB, particles can directly ionize the barrel APD photodetectors without striking a crystal, leading to a signal without scintillation light. These signals, which are known as spikes, occur at a rate that is directly proportional to the intensity of the LHC collisions, and the sFGVB is used to identify whether or not a signal is of this nature and should be suppressed from contributing to the trigger. The results of the L1 spike killer are dependent on the pedestal, or mean background noise read by ECAL electronics, as the sFGVB applies an energy threshold on a crystal-by-crystal basis. \\ \newline 
Pedestal measurements are performed regularly during the year, and an increase in the pedestal values has been seen in ECAL over time. Because of this, the effective threshold used by the sFGVB algorithm is reduced, as this pedestal has to be subtracted before the sFGVB acts. Therefore, the pedestals have to be updated periodically (twice in 2018) to maintain effective rejection of spikes. Fig 2 shows the effect of such an update, where an improvement in the L1 spike killer efficiency is seen:

\begin{figure}[H]
\centering
\includegraphics[height=3.1in]{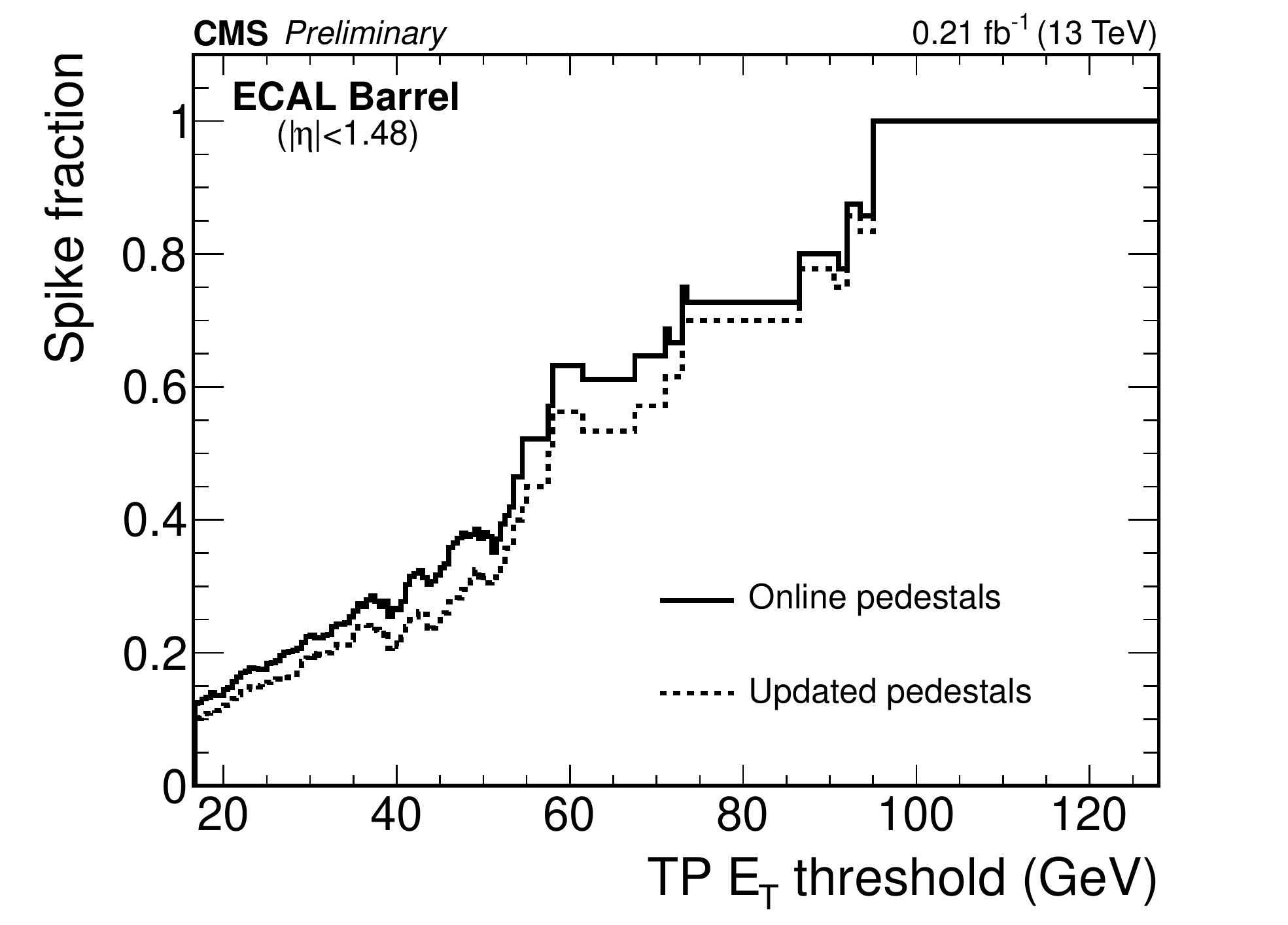}
\caption{L1 spike contamination in the ECAL TPs before and after pedestal update, using data recorded in July 2018.}
\end{figure}

\subsection{Pulse Shape Corrections}
Due to radiation received by ECAL over time, the pulse shape of the ECAL electronic signals changes with time. It has been seen that this is strongly dependent on the pseudorapidity of the crystals, with higher $\eta$ crystals receiving more radiation leading to more dramatically distorted pulses. If uncorrected, this can impact the bunch crossing assignment of the TPs, leading to substantial trigger inefficiencies. This is compensated for with a change in readout timing, which was performed three times in 2018. The resulting measurements shown in Fig 3 show an increasing time drift of signals in 2017, making regular corrections necessary.

\begin{figure}[H]
\centering
\includegraphics[height=3.1in]{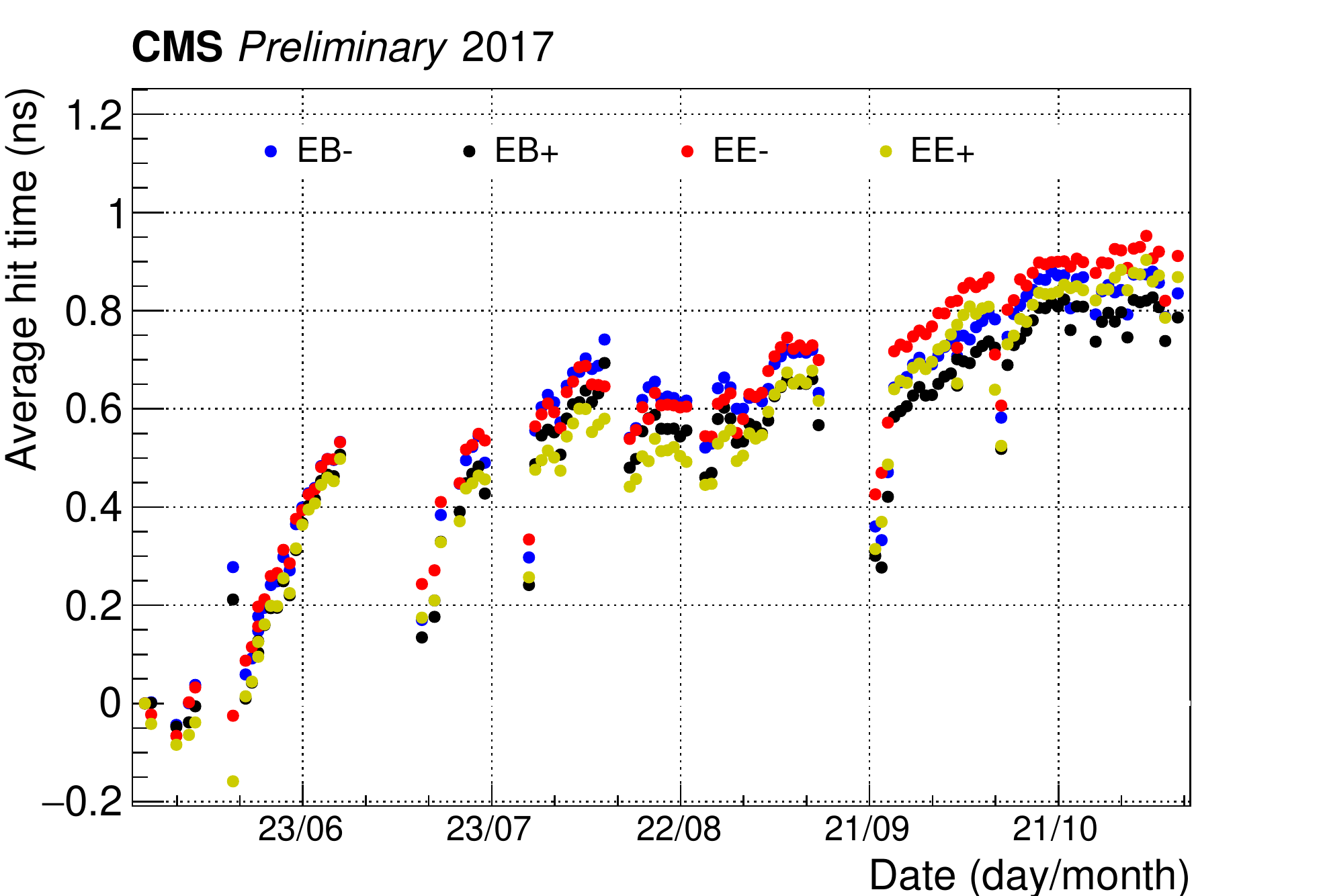}
\caption{ECAL pulse timing offset versus time during 2017.}
\end{figure}

\subsection{Run 2 Results}
As a result of the combination of these calibrations, in addition to calibrations made by the (hadronic calorimeter) HCAL detector and good overall detector performance, CMS maintained an excellent e/$\gamma$ trigger efficiency in Run 2, as seen in Fig 4:

\begin{figure}[H]
\centering
\includegraphics[height=3.75in]{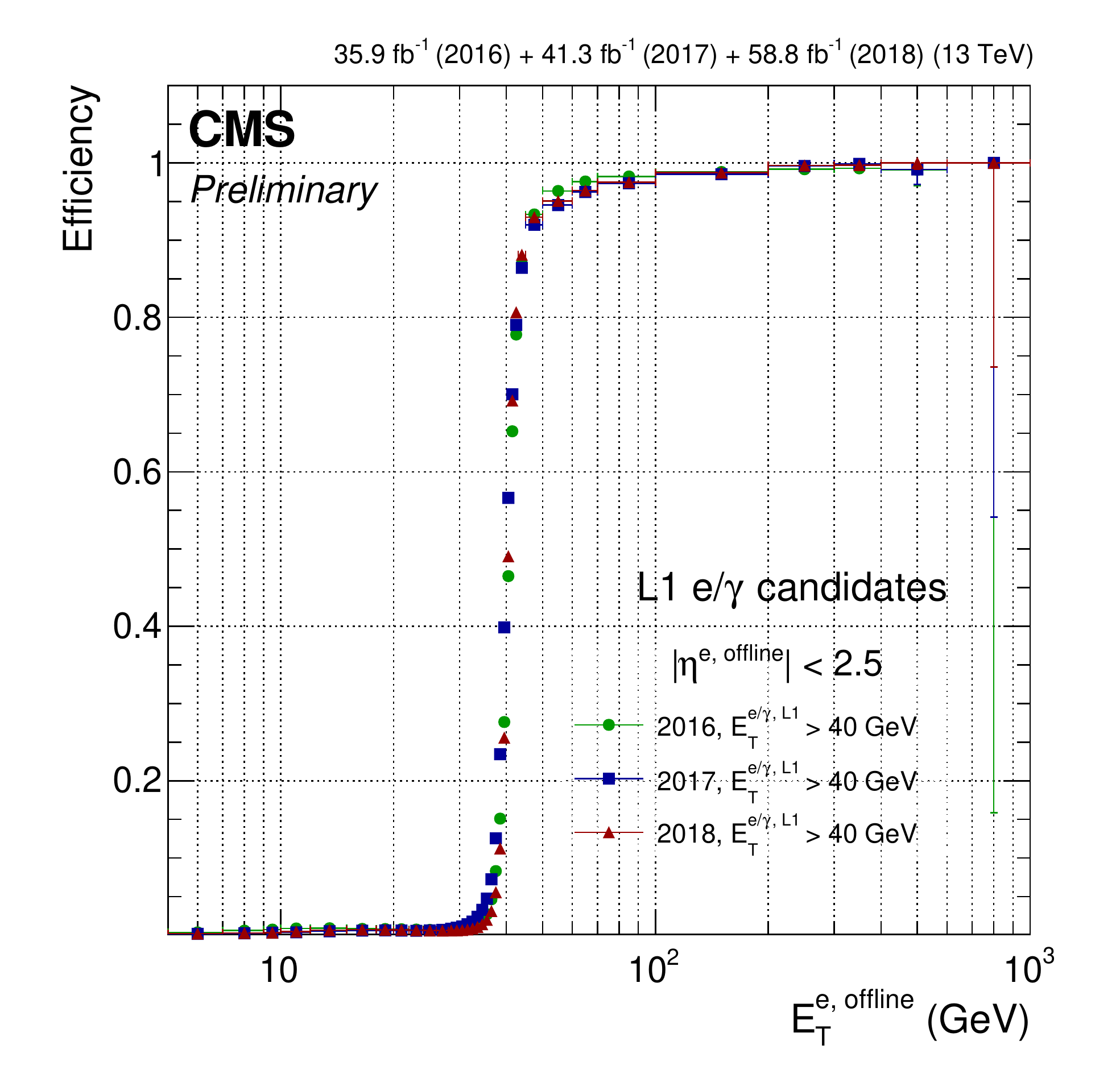}
\caption{L1 Turn-on curves for electron/photon candidates with $E_{T} >$ 40 GeV during Run 2, plotted as a function of offline reconstructed electron transverse energy.}
\end{figure}


\section{Optimization for Run 3}
To date, only one set of weights has been used for all EB signals, while a second set of weights was used for all EE signals. For Run 3 of the LHC, it is of interest to use the full capabilities of the on-detector electronics, which may allow for the setting of weights on a strip-by-strip basis, where a strip is a logical grouping of 5 adjacent crystals, the use of a second set of weights, and the use of an additional 6th weight. These possibilities are being tested on simulated and real data from ECAL in an attempt to test new L1 algorithms to improve ECAL trigger performance for Run 3. \\ \newline
While excellent L1 trigger efficiency was achieved during Run 2, it is of interest to maintain or even improve efficiency for Run 3 where increased detector ageing and pileup effects will be present. The methods with which this can be achieved arise from updating the current L1 amplitude weighs, and using unused features of the ECAL electronics.   

\subsection{Updating the Current Amplitude Weights}
Because ECAL pulse shapes have been changing over time, the weights applied at L1 are no longer ideal. This can be seen by applying Run 2 weights, which were derived from signal pulses from undamaged crystals, to signal pulses extracted from September 2018 CMS data, shown in Fig 5: 

\begin{figure}[H]
\centering
\includegraphics[height=4in]{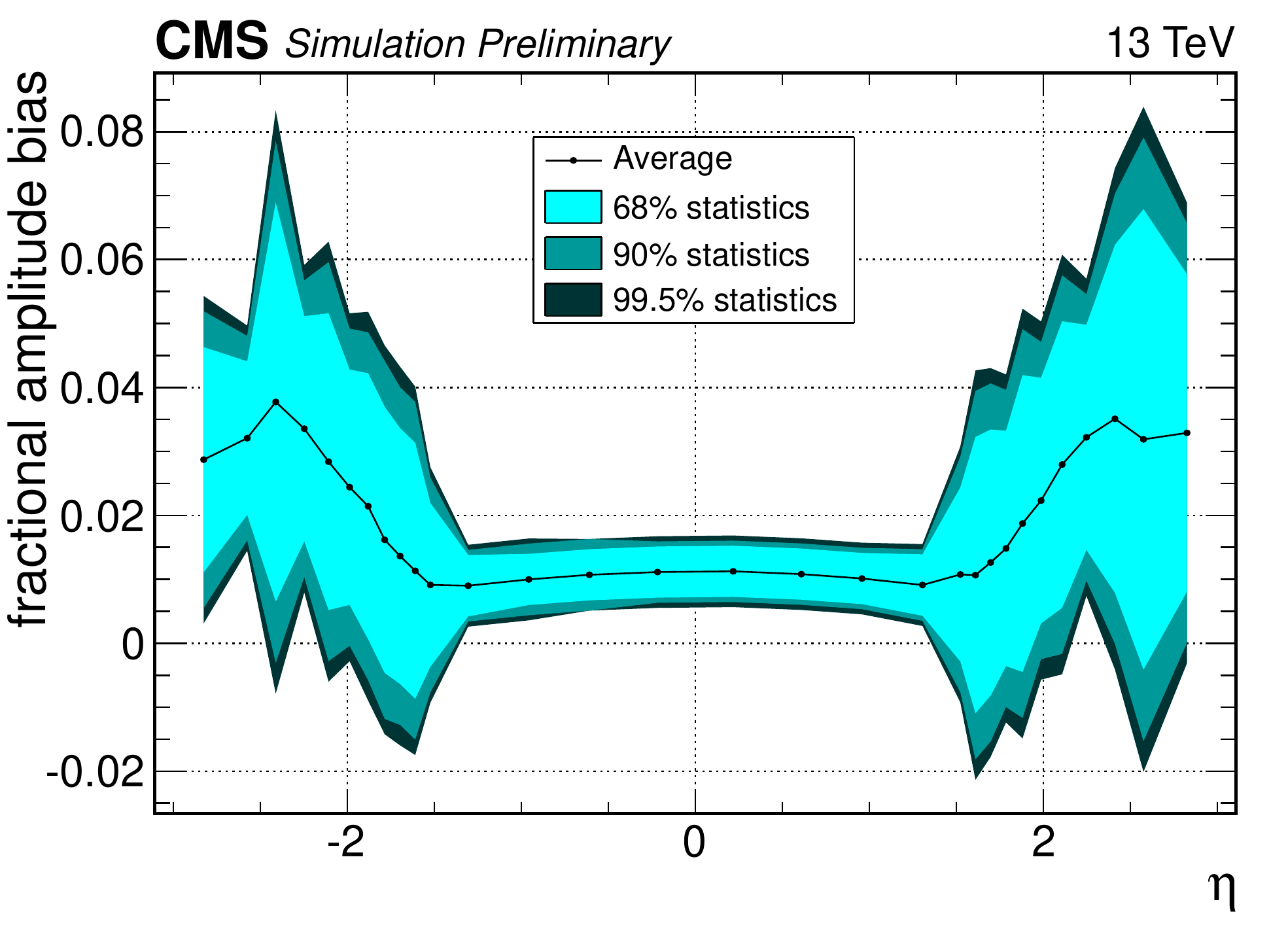}
\caption{Fractional amplitude bias vs. $\eta$ using Run 2 weights applied to waveforms derived from September 2018 data \cite{Plot}.}
\end{figure}
\noindent Amplitude bias is defined as the percent difference between reconstructed and true amplitude. Reconstructed amplitude is equal to the product of waveform samples and weights, and true amplitude is the height of the waveform. \\ \newline
There is an amplitude bias of $\approx 1\%$ in the EB, and a bias of up to 3\% in the higher $\eta$ regions in EE. This indicates that waveform shapes have changed, and more up to date weights should be used. \\ \newline
A simulation has been developed to model ECAL measurements for customized signal energy, pileup, and LHC bunch structure. This allows for the derivation of weights per strip of crystals with known data-taking conditions, and the creation of digitized ECAL waveform samples (digis) to simulate energy reconstruction. \\ \newline 
Updated amplitude weights have been derived from waveforms measured with September 2018 measured CMS data. A waveform was measured for each crystal, allowing for a set of weights to be derived for each crystal. When these weights are averaged over the EB and EE, updated sets of amplitude weights are obtained and applied to simulated ECAL digis. This distribution is labelled ``New (avg)" in Fig 6, where in this case only EE weights are used because the simulation is restricted to $2.3 < |\eta| < 3$. This results in an improved average bias compared to ``Current" weights, corresponding to the Run 2 weights. \\ \newline 
Per strip (PU=0) weights are derived from September 2018 measured waveforms without pileup in the LHC bunch structure, and averaged over each strip. \\ \newline 
When optimizing weights for different PU scenarios, the choice of pileup magnitude correlates with the energy of a pileup interaction at each bunch crossing, whose waveform is modelled by the average characteristic waveform of the ECAL strip of crystals. Weights are then derived using samples that include both signal and pileup. \\ \newline 
In all instances, the signal waveform shape is modelled by September 2018 measured waveforms whose height is set to the desired signal amplitude. \\ \newline 
When new weights are derived on a per strip basis and applied to simulated data, a decrease in bias average and spread is seen compared to that using the Run 2 weights, shown in Fig 6: 

\begin{figure}[H]
\centering
\includegraphics[height=4in]{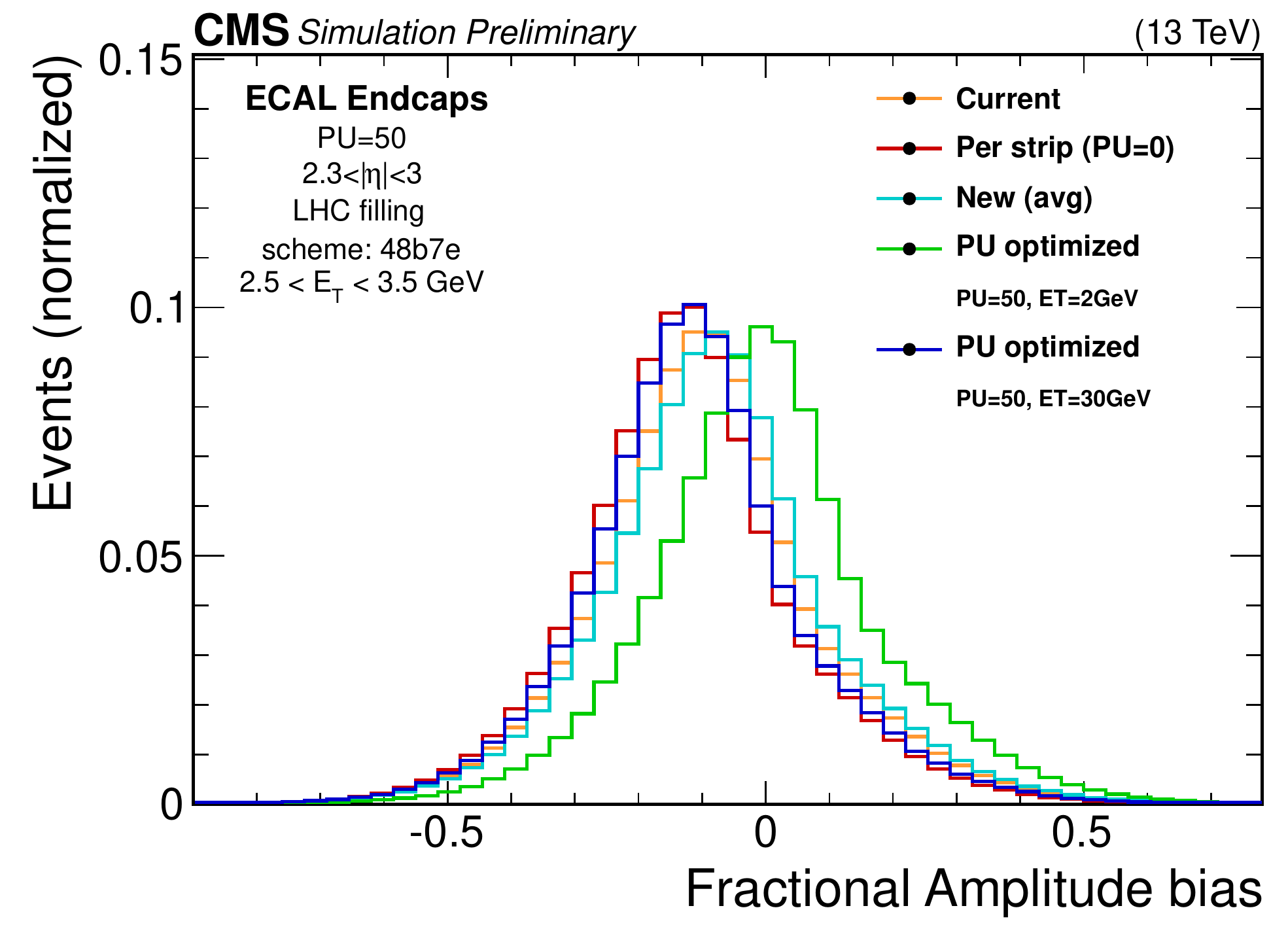}
\caption{Fractional amplitude bias using September 2018 waveforms, and multiple types of weights for a signal with 2.5 $<E_{T}<$ 3.5GeV in the range 2.3 $<|\eta|<$ 3.0 and for a simulated LHC pulse train with PU = 50}
\end{figure}
\noindent It can be seen that the weights perform better when they are optimized for the appropriate signal shapes, PU, and signal amplitudes. This indicates that for Run 3 weights can be optimized with a per strip granularity, and for specific pileup and signal amplitude scenarios.

\subsection{Second Set of Weights}
It is known that the FENIX chips to compute the TPs in ECAL on-detector electronics are capable of storing a second set of weights \cite{CMS}, which have never been used on-detector at ECAL. It is currently understood that an interplay between the results of these two weight sets applied to signals is possible within the electronics, allowing the use of two sets of information to decide to keep a TP rather than one. Current ideas for how to optimize and test a second set of weights are for large out of time pileup suppression, or possibly to improve the identification of spike-like signals in EB. 
\subsubsection{Timing Weights}
In addition to amplitude weights, timing jitter weights can be derived which, instead of measuring the amplitude of a waveform, measure the waveform's peak time relative to the expected peak time. Because spike and EM signals have different pulse shapes, it can be expected and has been shown that timing weights return different trends between EM shower like and spike-like signals, and that using timing information in the formation of ECAL TPs could further reject spikes at L1:
\begin{figure}[H]
\centering
\includegraphics[height=4in]{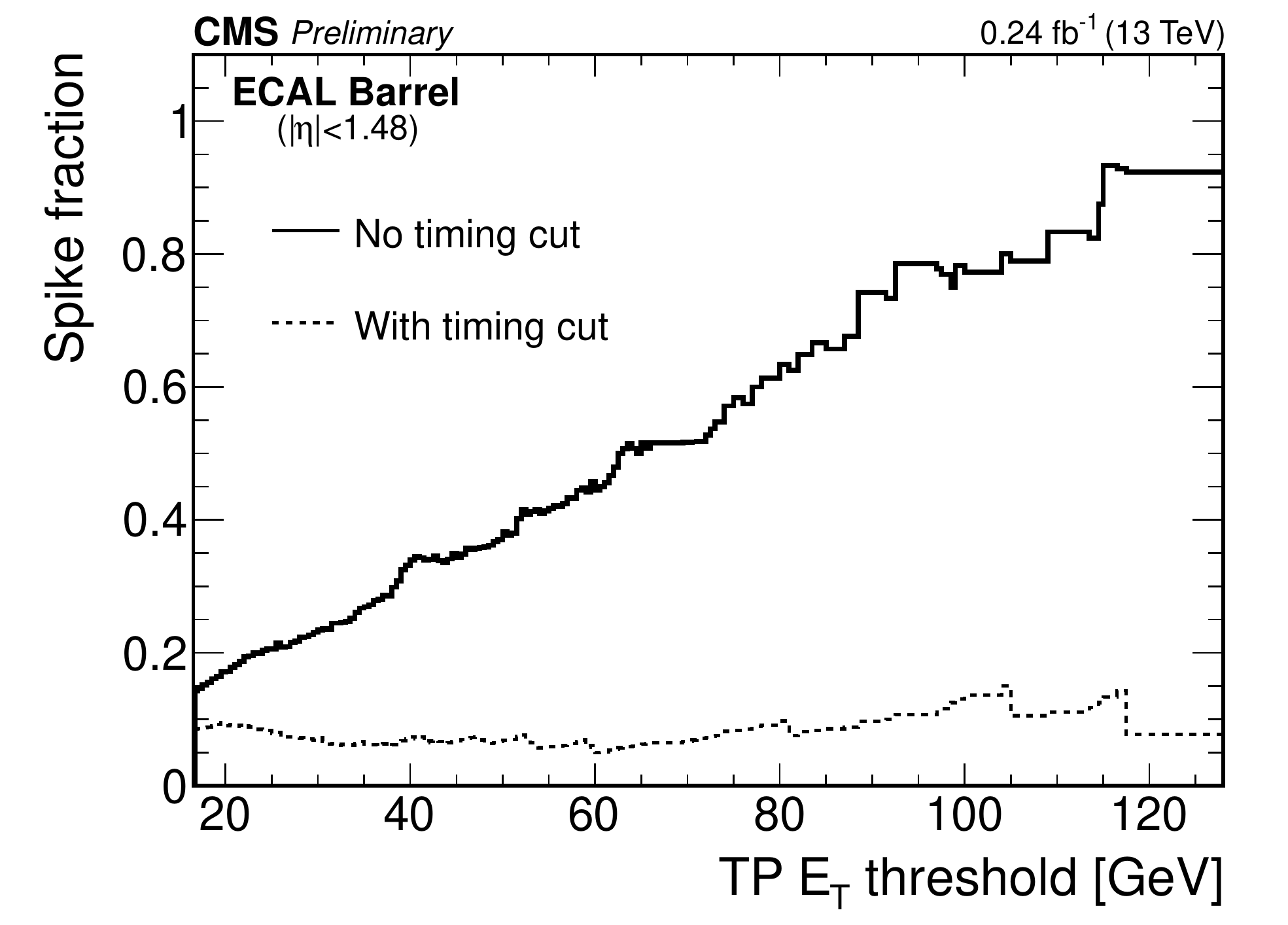}
\caption{Spike contamination before and after potential timing cut of -5 to 20 ns}
\end{figure}
\noindent It may be possible to use timing-optimized weights as a second set of weights in the FENIX in order to better reject events with spikes. This depends on the mechanism by which the two weight set outputs communicate, so their feasibility for use on ECAL must still be tested. While it has not yet been shown that a timing cut can be made at L1, Fig 7 shows that the potential outcome of making use of timing weight information can greatly reduce spike contamination. The feasibility of incorporating pulse timing information in the formation of ECAL TPs will be investigated for Run 3. 

\subsubsection{Out-of-Time Pileup Weights}
Studies are ongoing for the derivation of a second set of weights to be optimized to identify signals with large pulse distortions due to out of time PU. Such a set of  weights could in principle be used to suppress a signal if a large out-of-time pileup is identified. 

\subsubsection{6th Weight}
During Run 2, five amplitude weights were used to measure signal energy. The ECAL on-detector electronics have the capability to use an additional 6th weight. Studies are ongoing to see if the addition of a 6th weight can improve signal amplitude reconstruction, particularly in the forward regions of EE, where the effects of out-of-time PU are largest. 

\section{Conclusions}
ECAL online energy reconstruction is performed by applying fast algorithms to the signal pulses produced by the ECAL read-out electronics. Due to data taking challenges including high pileup, luminosity, and radiation, regular calibrations need to be made to maintain ECAL L1 trigger efficiency. \\ \newline 
During Run 2 of the LHC, regular recalibrations of the ECAL TPs were made including intercalibration, laser, pedestal, and pulse shape corrections in order to account for radiation induced changes in these quantities. These led to excellent ECAL TP energy scale stability and L1 trigger efficiency during Run 2. \\ \newline 
For Run 3, studies are being performed with the goal of maintaining and improving the ECAL L1 trigger performance. The methods for doing this include updating the Run 2 amplitude weights to be more optimal for new expected data-taking conditions, including signal pulse shapes, pileup and signal energy, and utilizing features of ECAL on-detector electronics that were unused during Run 2. These unused features include the possible use of a second set of L1 weights, and the inclusion of a 6th weight. \\ \newline 
It has been shown from simulation and data that updating amplitude weights and potentially using a second set of weights may improve amplitude reconstruction, and subsequently ECAL L1 trigger efficiency in Run 3. 

\section*{Acknowledgements}
Special thanks to Toyoko Orimoto, David Petyt, William Richard Smith, Davide Valsecchi and the CMS collaboration.

\end{document}